\documentclass{aa}  
\usepackage{graphicx}
\usepackage{multirow}
\usepackage[dvipsnames]{xcolor}
\usepackage{txfonts }
\usepackage{natbib  }
\usepackage{hyperref}
\newcommand{\RA}[3]{$#1^\mathrm{h} #2^\mathrm{m} #3^\mathrm{s}$}
\newcommand{\Dec}[3]{$#1^\circ     #2^\prime     #3^{\prime\prime}$}
\newcommand{\Jyb}{\textrm{Jy~beam}$^{-1}$}
\newcommand{\kms}{km~s$^{-1}$}
\newcommand{\amin}{$^{\prime}$}
\newcommand{\asec}{$^{\prime\prime}$}
\newcommand{\degree}{$^\circ$}
\newcommand{\dms}[3]{$#1^{#2}\!\!.#3$}
\newcommand{\cm}[1]{$\mathrm{cm}^{#1}$}
\newcommand{\Msun}{$\mathrm{M_\sun}$}

\def\kes{{Kes~17}}

\def\SF305{{G305.136+0.068}}
\def\HII304{{G304.465--00.023}}
\def\FGLsource{{4FGL~J1305.5--6241}}

\def\Fermi{{\it Fermi}}
\begin{document}

\title{The SNR Kes 17-ISM interaction: a fresh view from radio and $\gamma$ rays}
\author{Authors: TBD}
\author{L. Sup\'an\inst{1}
   \and G. Castelletti\inst{1}
   \and A. Lemi\`ere\inst{2}}
\institute{Instituto de Astronomía y Física del Espacio (IAFE), CONICET-Universidad de Buenos Aires, 1428, Buenos Aires, Argentina\\
\email{lsupan@iafe.uba.ar}
\and Universit\'e de Paris, CNRS, Astroparticule et Cosmologie, F-75013 Paris, France}

\date{Received DD Month 2023 / accepted DD Month 2023}
\abstract
{This paper presents a comprehensive analysis of the Galactic supernova remnant (SNR) {\kes} (G304.6+0.1) with a focus on its radio synchrotron emission, environs, and the factors contributing to the observed $\gamma$ rays. 
The fitting to the firstly-obtained integrated radio continuum spectrum spanning from 88 to 8800~MHz yields an index $\alpha=-0.488\pm0.023$ ($S_{\nu}\propto\nu^{\alpha}$), indicative of a linear particle acceleration process at the shock front of the remnant. 
Accounting for the SNR radio shell size, the distribution of atomic hydrogen ($n_{\mathrm{H}} \sim 10$~\cm{-3}), and assuming the SNR is in the Sedov-Taylor stage of its evolution, we estimate the remnant's age to be roughly 11~kyr. This result falls at the lower end of the wide range ($\sim$2-64~kyr) derived from previous analyses of the diffuse X-ray emission interior to the remnant.
Furthermore, we used \element[][12]{CO} and \element[][13]{CO} ($J$=1-0)
emission-line data  as a proxy for molecular hydrogen and provided the first evidence that the eastern shell of {\kes} is engulfing a molecular enhancement in the surrounding gas, with an average mass $4.2 \times 10^4$~{\Msun} and density $n \sim 300$~\cm{-3}. 
Towards the western boundary of {\kes} there are not signatures  of carbon monoxide emissions above 3$\sigma$, despite previously reported infrared observations have revealed shocked molecular gas at that  location. 
This suggests the existence of a CO-dark interacting molecular gas, a phenomenon also recorded in other Galactic SNRs (e.g. CTB~37A and RX~J1713.7--3946) revealing itself both in the infrared and $\gamma$-ray domains. 
Additionally, by analysing  $\sim$14.5~yr of continuously collected  data from the Large Area Telescope on board the satellite {\Fermi}, we determined that  
the best-fit power-law photon index for  the 0.3-300~GeV $\gamma$-ray emission from the {\kes} region is $\Gamma=2.39\pm0.04^{+0.063}_{-0.114}$ ($\pm$stat $\pm$syst) in agreement with prior studies. 
The energy flux turns out to be $(2.98 \pm 0.14) \times 10^{-11}$~erg~cm$^{-2}$~s$^{-1}$ 
implying a luminosity $(2.22 \pm 0.45) \times 10^{35}$~erg~s$^{-1}$ at $\sim 8$~kpc. 
Finally, we successfully modelled the multiwavelength spectral energy distribution by incorporating the 
radio synchrotron spectrum and the new measurements of GeV $\gamma$-rays. Our analysis indicates that the observed $\gamma$-ray flux most likely originates from the interaction of {\kes} with western ``dark'' CO zone with a proton density $n_{\mathrm{p}} \sim$400~cm$^{-3}$.}

\titlerunning{A fresh view of SNR Kes~17-ISM interaction}
\authorrunning{Sup\'an et al.}
\maketitle

\section{Introduction}
\label{intro}
Supernova remnants (SNRs) are captivating objects that cause  a long-lasting impact on the Galactic ecosystem, leaving distinct imprints that can be observed across the entire electromagnetic spectrum.
This paper, focused on the source {\kes} (G304.6+0.1), is part of a series of articles conducted by the author team, dedicated to investigating the association between radio and $\gamma$-ray emissions in remnants of stellar explosions. Previous studies in this series were devoted to  G338.3$-$0.0 \citep{Supan+16_G338}, Kes~41 \citep{supan2018,Supan+18_kes41-MC}, and G46.8$-$0.0 \citep{Supan+22_G46}, all of them middle-aged  $\gamma$-ray emitting SNRs interacting with their ambient medium.

The initial observations of {\kes} at radio wavelengths were conducted in the 1970s at 408 and 5000~MHz using the Molonglo and Parkes single-dish telescopes \citep{Shaver+Goss-70_I,Shaver+Goss-70_II,Milne+Dickel-75}. 
The first distance estimate for the remnant was approximately 6~kpc, derived using the uncertain $\Sigma$-$D$ relation \citep{Shaver+Goss-70_III}. 
Later on, \citet{Caswell+75} established a lower limit of 
9.7~kpc based on absorption features in the low-resolution neutral hydrogen (\ion{H}{I}) spectrum of gas clouds along the line-of-sight. This lower limit remained unchanged for nearly five decades, until a recent reanalysis by \citet{Ranasinghe+Leahy-22_distances}, which yielded a kinematic distance $7.9 \pm 0.6$~kpc to the remnant.

{\kes} was extensively studied in  the X-ray domain using data obtained with XMM-\it Newton\rm,  \it Suzaku\rm, and \it ASCA \rm satellites \citep{Combi+10,Gok+Sezer-12,Gelfand+13,Pannuti+14,Washino+16}. 
According to the observed properties of the X-ray emitting gas, \citet{Combi+10} proposed that the source belongs to the mixed-morphology (MM) type of SNRs, characterised by  a shell-like morphology in radio wavelengths and a filled-centre composition in X rays. Additionally, they suggested the presence of a nonthermal component in the northern, central, and southern regions of the SNR shock front. 
However, subsequent studies with improved statistics have raised doubts about this possibility and concluded that the X-ray spectrum is dominated by thermal emission \citep{Gok+Sezer-12,Gelfand+13,Washino+16}. 
At the high-energy end of the spectrum, {\kes} has been linked to a GeV source detected by the \it Fermi \rm Large Area Telescope (LAT) \citep{Wu+11,Gelfand+13}. No counterpart at TeV energies has been reported  so far.
Concerning the SNR environment, the first hint of  interaction came from low-resolution observations of the 1720-MHz maser-line of hydroxyl (OH) \citep{Frail+96}. 
A significant breakthrough in linking {\kes} to the interstellar matter occurred through near-infrared (near-IR) spectroscopic studies, which were crucial in firmly determining the location and characteristics of the  shocked H$_{2}$ gas \citep{Lee+11_FIR}.

This paper is organised as follows: In Sect.~\ref{radio-spectrum-age} we present the first analysis of the radio continuum spectrum for {\kes}. Adopting a standard evolution model we also derived the SNR's age by means of the radio size of the  expanding forward shock and 
21~cm spectral-line observations of the \ion{H}{I} gas. 
Sect.~\ref{CO} focuses on investigating the morphology and kinematics of the molecular gas emitting in CO lines.  This represents the first study of the CO gas in the region of {\kes}.
In Sect.~\ref{gamma}, we provide an updated analysis of the {\Fermi}-LAT data covering 14.5~yr. We also investigate the emission mechanism responsible for the high-energy flux through a broadband modelling that incorporates the revisited measurements at radio and GeV $\gamma$-ray energies. Our findings are summarised in Sect.~\ref{summary}.

\section{SNR~{\kes} in radio wavelengths}
\label{radio-spectrum-age}
\subsection{Morphology and spectrum}
\label{radio-spectrum}
The radio remnant {\kes} is characterised by non uniform emission from a complete, albeit  irregular, shell structure with an average size of  $\sim 7^{\prime}$. 
This can be observed in the 843~MHz image from the Molonglo Sky Survey (SUMSS, HPBW = 45$^{\prime\prime} \times 50^{\prime\prime}$)\footnote{\url{http://www.physics.usyd.edu.au/sifa/Main/SUMSS}, \citet{Bock+99}} included in the inset of Fig.~\ref{CO-grid}{\it a}. 
The surface brightness of the remnant is $\sim 4.3 \times 10^{-20}$~W~m$^{-2}$~Hz$^{-1}$ at 843~MHz, while that of the brightest elongated feature ($\sim$\dms{4}{\prime}{2}~$\times$~\dms{1}{\prime}{2} in size) along the southern periphery is $\sim 0.11 \times 10^{-20}$~W~m$^{-2}$~Hz$^{-1}$
An arc of enhanced synchrotron emission is noticed in the northwest region of the remnant, near \RA{13}{05}{30}, \Dec{-62}{41}{00}. 
This arc has a size of approximately $\sim$\dms{1}{\prime}{3}~$\times$~\dms{2}{\prime}{5} and coincides with a distinctive bend in the shock front of {\kes}. 
Bright continuum and line emission in the IR wavebands, as reported by \citet{Lee+11_FIR}, accompanies the radio synchrotron emission along this edge of {\kes}. 
Another noteworthy feature is an indentation towards the east of the radio shell, at approximately \RA{13}{06}{05}, \Dec{-62}{41}{10}, possibly indicating that the SNR shock is wrapping around an external inhomogeneity. 
The structure of the surrounding matter, revealed for the first time by the  analysis of the CO gas is discussed in Sect.~\ref{CO}.

To construct the global spectrum of the radio continuum emission of {\kes}, we compiled flux density estimates from the literature as well as new fluxes that we measured from publicly available radio surveys. 
For frequencies below $\sim$160~MHz, the lowest  at which {\kes} has been detected to date, we used the Galactic and Extragalactic All-sky Murchison Widefield Array Survey (GLEAM, \citealt{hurley19})%
\footnote{\url{https://www.mwatelescope.org/science/galactic-science/gleam/}}. 
We also used the Southern Galactic Plane Survey (SGPS, \citealt{McClure+05})%
\footnote{\url{https://www.atnf.csiro.au/research/HI/sgps/queryForm.html}}at 1420~MHz  and 
the S-band Polarisation All Sky Survey (S-PASS, \citealt{Carretti+19})%
\footnote{\url{https://sites.google.com/inaf.it/spass}} at 2303~MHz. 
Flux measurements with an error greater than 20\%  were discarded from the analysis. 
For the remaining data, if information on the primary calibrator was available, they were brought to the absolute flux scale presented by \citet{perley+17}. This flux scale is valid for the entire range of compiled frequencies in our analysis (88-8800~MHz)  and has an accuracy of $\sim$3\%. 
The set of data points was fitted using the simple power-law model $S_\nu \propto \nu^\alpha$, where $S_\nu$ represents the integrated flux at frequency $\nu$ and $\alpha$ is the radio spectral index. 
During the fitting process, flux measurements were rejected if their dispersion with respect to the model was greater than 2$\sigma$ of the best-fit values. 

\begin{figure}[h!]
  \centering
  \includegraphics[width=0.47\textwidth]{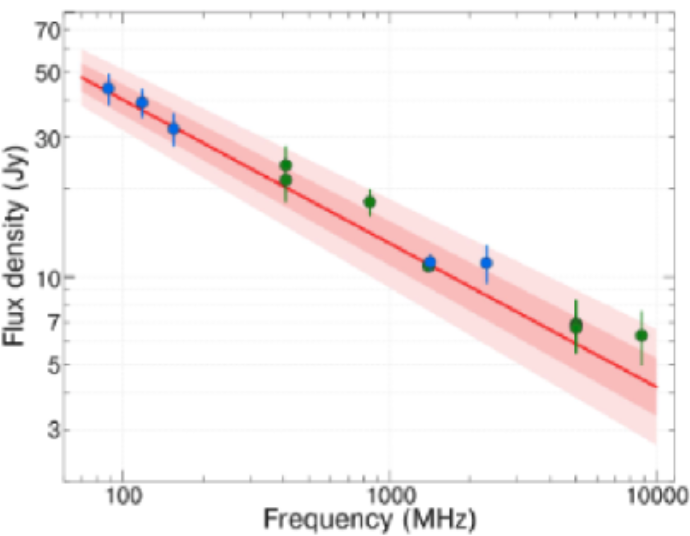}
  \caption{Spectrum of the continuum emission from SNR {\kes} at radio frequencies, constructed with the fluxes in Table~\ref{fluxes-table}. 
  Blue data points denote our new measurements from public survey images, while the green ones correspond to literature flux estimates. 
  The straight line represents the best fit with a power-law model in the form $S_\nu \propto \nu^\alpha$, which yields a spectral index value $\alpha = -0.488 \pm 0.023$. 
  Darker and lighter pink-shaded regions around the straight line denote a variation in the fitted spectral parameters of 1 and 2$\sigma$, respectively.}
 \label{kes17_radio-spectrum}
\end{figure}

The final dataset is reported in Table~\ref{fluxes-table}. 
It constitutes the most complete compilation of radio flux measurements conducted for {\kes} to date. 
Figure~\ref{kes17_radio-spectrum} displays the integrated radio continuum spectrum for this SNR, with our new flux density determinations represented by blue filled circles. A weighted least-squares fit was applied to the data points resulting a spectral index $\alpha =-0.488 \pm 0.023$. This value is flatter than the previous measurement ($\alpha \simeq -0.54$) reported by \citet{Shaver+Goss-70_III}, which was based on flux estimates at only 408 and 5000~MHz. 
The synchrotron radiation spectrum we derived for {\kes} is consistent with 
electrons being accelerated via a first-order Fermi mechanism \citep{Onic+13}. 
Regarding the spectral shape, the straight distribution of flux densities at low radio frequencies (below 100~MHz) indicates that if ionised gas exists, whether it is located either co-spatially or coincidentally intersecting {\kes} along the line of sight, the free-free absorption it produces does not have an impact on the integrated continuum spectrum of the remnant. Low-frequency turnovers caused by free-free absorption by ionised gas in \ion{H}{II} regions (or in their associated lower-density envelopes), as well as at the interface between an ionising shock and its immediate environment, have been measured in the spectra of some SNRs (e.g. Kes~67, Kes~75, W41, Kes~73, 3C~396, and W49B, \citealt{castelletti21}). 
For {\kes}, determining whether its forward shock ionises or not the western region where SNR's interaction with dense gas has been proved in the infrared \citep{Lee+11_FIR}, could potentially be illuminated by improved sensitivity and resolution radio observations, especially at the low-frequency portion of the spectrum.

\begin{table}[h!]
\small\centering
\caption{Integrated flux densities over the full extent of {\kes} used to construct the radio continuum spectrum of the remnant shown in Fig.~\ref{kes17_radio-spectrum}.} 
\begin{tabular}{ccl}
\hline\hline
  Frequency  &  Integrated flux  &  \multirow{2}{*}{Reference} \\
    (MHz)    &     (Jy)          &                             \\\hline
       88  &  43.9 $\pm$ 5.2  &  This work (GLEAM)          \\
      118  &  39.3 $\pm$ 4.2  &  This work (GLEAM)          \\
      155  &  32.0 $\pm$ 4.0  &  This work (GLEAM)          \\
      408  &  21.4 $\pm$ 2.2  &  \citet{Shaver+Goss-70_II}  \\
      408  &  24.0 $\pm$ 3.7  &  \citet{Kesteven-68}        \\
      408  &  21.4 $\pm$ 3.3  &  \citet{Green-74}           \\
      843  &  18.0 $\pm$ 1.8 \tablefootmark{$\dag$} &  \citet{MOST-catalogue} \\
     1400  &  10.9 $\pm$ 0.14\tablefootmark{$\dag$} &  \citet{Gelfand+13}     \\
     1420  &  11.2 $\pm$ 0.7 \tablefootmark{$\dag$} &  This work (SGPS)       \\
     2303  &  11.1 $\pm$ 1.6 \tablefootmark{$\dag$} &  This work (S-PASS)     \\
     5000  &   6.7 $\pm$ 0.7  &  \citet{Shaver+Goss-70_III} \\
     5000  &   6.8 $\pm$ 0.7  &  \citet{Milne-69}           \\
     5000  &   6.9 $\pm$ 1.4  &  \citet{Milne+Dickel-75}    \\
     8800  &   6.3 $\pm$ 1.3  &  \citet{Dickel+73_8.8GHz}   \\\hline
 \label{fluxes-table}
\end{tabular}
\tablefoot{\tablefoottext{$\dag$}
{Measurements not corrected for the absolute flux scale of \citet{perley+17} due to missing information on the primary flux calibration.}}
\end{table}

\subsection{SNR's age and neutral gas properties from radio data} 
\label{age-NH} 
Estimating the dynamical age of supernova remnants involves indirect methods due to the inability to measure it directly. 
So far, all age estimates for {\kes} have been derived on the basis of 
spectral fitting parameters to the cold and low density X-ray emitting gas inner to the radio shell. 
However, there are large differences in the estimated ages, with values ranging from 2.3~kyr to as high as 64~kyr depending on factors such as ionisation timescales, $\tau$, and electron densities, $n_{\mathrm{e}}$ ($\tau \sim$ 1-3 $\times$ $10^{12}$~cm$^{-3}$~s, $n_{\mathrm{e}} \sim$ 0.4-2.3~cm$^{-3}$, \citealt{Pannuti+14}, and references therein). Besides, according to the proposal that {\kes} is a member of the MM  SNRs group  and considering a thermal conduction model, \citet{Gelfand+13} derived an age range from 2 to 5~kyr. 
Additionally, they determined an upper age limit of 40~kyr by assuming that clump evaporation into the inter-cloud medium is primarily responsible for the observed X-ray emission. 

In this section, we employ a standard evolution model and examine continuum and line emissions at centimetre wavelengths to estimate the age of {\kes}. 
The relative intermediate extent of {\kes}'s shock front (as illustrated in the inset of  Fig.~\ref{CO-grid}a) compared to other SNRs discovered in the Galaxy, coupled with the absence of optical signatures of radiative shocks in observations from surveys like the SuperCOSMOS H-alpha Survey (SHS, \citealt{parker+05})%
\footnote{\url{http://www-wfau.roe.ac.uk/sss/halpha/index.html}} 
or the STScI Digitized Sky Survey (DSS),%
\footnote{\url{https://archive.stsci.edu/cgi-bin/dss_form}.} 
lends support to the hypothesis that {\kes} is in the Sedov expansion stage of its evolution. 
Based on this picture, the time elapsed since the explosion can be estimated via the relation \citep{cox1972}: 
\begin{equation}
t_{\mathrm{SNR}}\simeq
\left(\frac{R_{\mathrm{S}}}{12.9}\right)^{5/2} \, 
\left(\frac{n_{0}}{\epsilon_{0}}\right)^{1/2} \, 10^{4}\, \mathrm{yr},   
\label{age}
\end{equation}

\noindent
where $R_{\mathrm{S}}$ is the shock radius at present (in pc), $n_{0}$ is the ambient interstellar density (in cm$^{-3}$), and $\epsilon_{0}$ is the initial explosion energy (in units of $0.75 \times 10^{51}$~erg). 
The radius, measured in the radio continuum image of {\kes} at 843~MHz, is $\sim$$3.5^{\prime}$ or $\sim$8~pc according to the revisited kinematic distance 
$7.9 \pm 0.6$~kpc to the SNR obtained by \citet{Ranasinghe+Leahy-22_distances} from neutral hydrogen \ion{H}{I} absorption features. 
For the ambient interstellar density, we considered it is well represented by the neutral hydrogen gas density, and estimated it via $n_{0}=N_{\mathrm{H}}/L$, the ratio of the
hydrogen column density over its depth in the line of sight in the region of the SNR. 
To calculate $N_{\mathrm{H}}$ we considered the 21~cm line emission of \ion{H}{I} from the SGPS data. 
Our focus was to identify any sign of neutral gas that could have been swept up by the SNR shock or by the stellar winds of the progenitor star. 
If we detect accumulation of \ion{H}{I} around the radio continuum boundary of the remnant, it can provide us with a rough estimate of the pre-shock medium density under the assumption that the accumulated atomic gas was uniformly distributed inside the volume of the \ion{H}{I} shell before the stellar explosion.
We did not find, however, any neighbouring structure of neutral atomic gas that could be feasibly associated with {\kes}. 
Moreover, an inspection of the HI datacube shows the remnant in absorption in the complete velocity range from $\sim$0~{\kms} to the tangent point velocity ($v_{\mathrm{TP}}\simeq -42$~{\kms}, according to the Galactic rotation curve of \citealt{Reid+14}). 
Therefore, we simply hypothesised that $n_{\mathrm{H}}$ could be well represented by  the mean density value measured in circular test-areas of radius $\simeq 8^{\prime}$  distributed around the remnant (we tested different values and determined that our result remains consistent, within uncertainties, regardless of the size chosen). 
Under the common assumption that the \ion{H}{I} emission is optically thin, after subtracting an appropriate mean background level to each \ion{H}{I} velocity channel, the mean column density around {\kes}, calculated by integrating the \ion{H}{I} emission between $-31$ and $-14$~{\kms} is $N_{\mathrm{H}} \approx 8 \times 10^{20}$~\cm{-2}. 
This velocity interval is in accord with the interstellar molecular matter traced in CO associated in space and velocity with the SNR (further analysis of this topic is provided in Sect.~\ref{CO}). 
According to all the assumptions we made previously, our analysis produces a number density in the \ion{H}{I} ambient environment $n_{0} \approx 7$~\cm{-3}, larger than the typical 
value $\sim$1 cm$^{-3}$ averaged over the cold, warm, and hot gas phases of the ISM \citep{mckee1977}. 
Therefore, using Eq.~\ref{age} and adopting the value $\sim 4 \times 10^{50}$~erg for the energy released in the SN event, as derived by \citet{leahy2020} incorporating both uniform ISM and stellar wind SNR evolutionary models, we estimated that {\kes} is approximately 11~kyr old. Notably, when using the kinetic energy $10^{51}$~erg for a canonical SN  the age decreases to roughly 7~kyr,  not critically different within the uncertainties from our calculation. 

We are aware that our approach for the {\kes}'s age provides a first-order approximation, since 
i) it assumes that  the SNR is in the Sedov stage of its evolution, 
ii) the mean number density of atomic hydrogen, as measured from HI data, represents an upper limit because some of the \ion{H}{I} may be unrelated gas located behind the SNR, 
and 
iii) our result ignores  the possibility of the remnant evolving in an inhomogeneous ambient medium.

\section{The molecular environment of {\kes}}
\label{CO}
The properties of the molecular gas in the region of {\kes}, as traced by the emission from carbon monoxide (CO), did not receive attention in previous works. Dense interstellar material interacting with the western shock front of the SNR was only revealed 
in infrared wavebands \citep{Lee+11_FIR}. 
Here, we present the main results of the first study towards {\kes} carried out by using both \element[][12]{CO} and \element[][13]{CO} emission data in their rotational transition $J$ = 1-0. The datacubes for both species were extracted from the Three-mm Ultimate Mopra Milky Way Survey (ThrUMMS, \citealt{Barnes+15}).%
\footnote{\url{https://users.astro.ufl.edu/~peterb/research/thrumms/}} 
The spatial and spectral resolutions are 72{\asec} and $\sim$0.35~{\kms}, respectively, with sensitivities $\sim$ 1~K each. 

\begin{figure*}[h!]
 \centering
\includegraphics[width=0.75\textwidth]{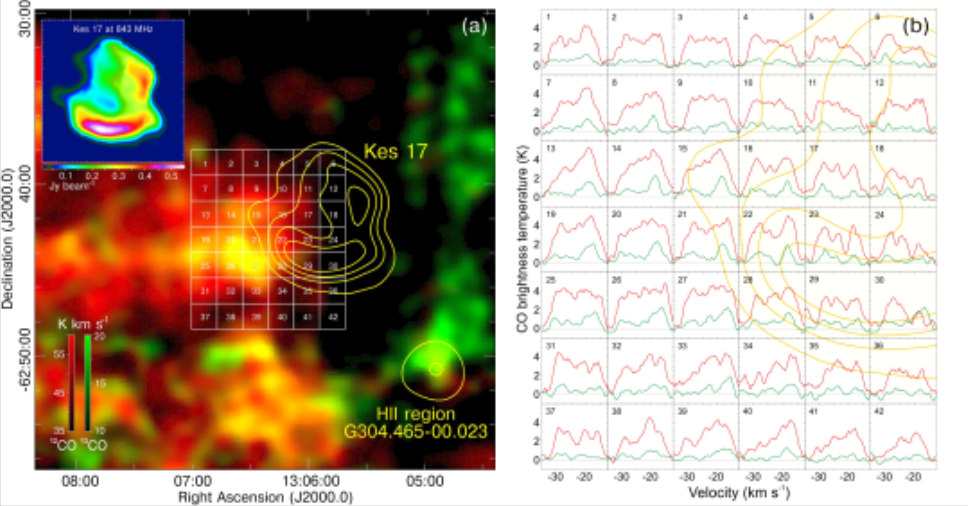}
 \caption{Spatial and spectral distribution of the CO gas towards {\kes} from the ThrUMMS \citep{Barnes+15}.  
 ({\it a}) Colour-coded image of the \element[][12]{CO} (in red) and the and \element[][13]{CO} (in green) $J=$1-0  line emissions, integrated from $-31$ to $-14$~{\kms}. 
 Yellow regions are areas where both CO isotopologue  emissions overlap. 
 Contours (levels: 0.034, 0.35, 0.75, and 1.2~{m\Jyb}, with a beam smoothed to the 80{\asec} spatial resolution of the CO data) delineate the 843~MHz-continuum radiation from {\kes}. 
 For reference, the inset in the upper left corner displays the structure of the SNR shell as observed by the SUMSS  (45{\asec} resolution). 
 The overlaid grid consists of the series of boxes (\dms{1}{\prime}{5} in size) used for analysing the kinematics of the CO gas.
 The \ion{H}{II} region G304.465--00.023 in the field is also labelled \citep{Urquhart+22}.
 ({\it b}) Collection of \element[][12]{CO} (in red) and \element[][13]{CO} (in green) $J=$1-0 spectra extracted from the boxes, numbered from 1 to 42, in panel ({\it a}). 
 The original datasets were convolved to a resolution of 80{\asec} to reduce the graininess.
 The yellow contours superimposed on the spectra correspond to the radio continuum emission from {\kes}.}
 \label{CO-grid}
\end{figure*}

\begin{figure*}[h!]
 \centering
 \includegraphics[width=\textwidth]{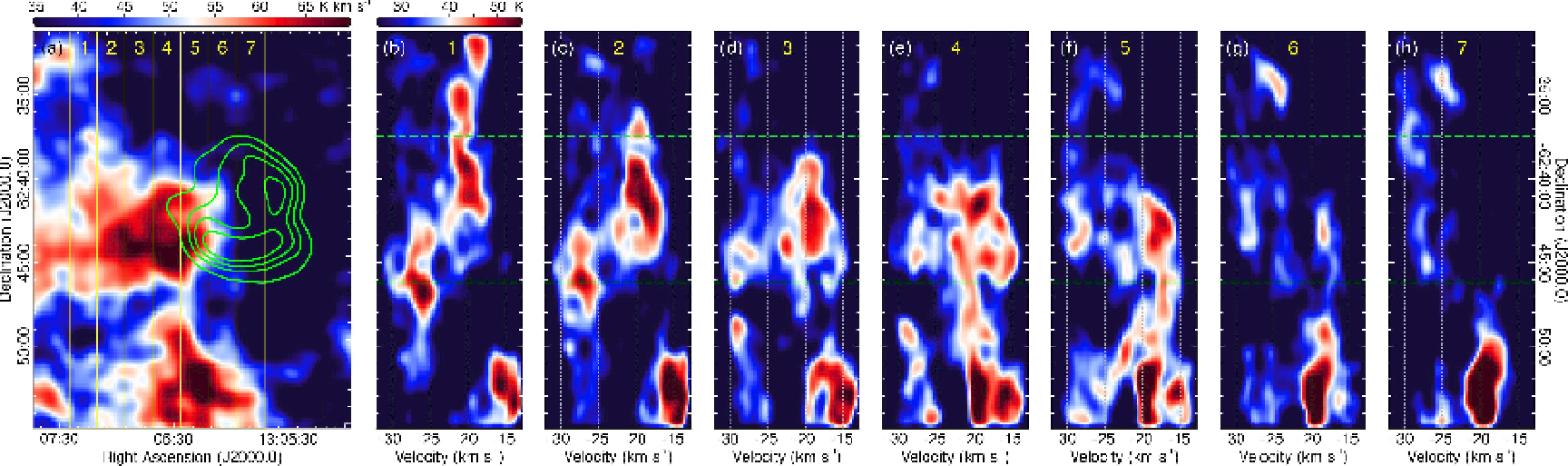}
 \caption{({\it a}) Regions, numbered from 1 to 7, used to construct the position-velocity ($p$-$v$) diagrams for the eastern cloud associated with {\kes},  shown in panels {\it b}-{\it h} . Colours indicate the spatial distribution of the \element[][12]{CO} $J=$1-0 gas  integrated in the ($-31$, $-14$)~{\kms} velocity range as in Fig.~\ref{CO-grid}, while contours trace the 843-MHz radio continuum emission from {\kes}. 
 ({\it b-h}) $p$-$v$ diagrams derived from the \element[][12]{CO} line emission depicted in panel ({\it a}). 
 They were constructed by integrating in the 100{\asec} R.A. interval indicated by the regions numbered from 1 to 7 in panel ({\it a}). 
 Horizontal dashed lines mark the extent of {\kes} in the Declination dimension.
 The colour representation is the same for all $p$-$v$ diagrams.}
 \label{CO-PV}
\end{figure*}

After carefully inspecting the CO data cubes throughout their velocity ranges $(-65,+55)$~{\kms}, we only found molecular structures in projected correspondence with the radio continuum emission from the SNR's shell in two intervals, with velocity ranges from $\sim -45$ to $-37$~{\kms} and $\sim -31$ to $-14$~{\kms}, respectively. 
The CO structure in the first range peaks at $-41$~{\kms} and in the plane of the sky lies towards the eastern border of {\kes} with an angular size of approximately 8{\amin}~$\times$~4{\amin} (in the south-north and east-west directions, respectively). We assigned to this molecular material a distance $\simeq$ 5~kpc, as its velocity $\simeq$ $-41$~{\kms} is largely consistent with that of the tangent point in the direction of {\kes}. 
Consequently, we discard this cloud as possibly associated with {\kes} ($d_\mathrm{SNR}$ $\simeq$ 8~kpc, \citealt{Ranasinghe+Leahy-22_distances}), and it will not be analysed in the following. 
The emission from the second velocity component originates from a cloud at the eastern part of the SNR shell. 
The distribution of integrated \element[][12]{CO} and \element[][13]{CO} emissions are presented in the colour-coded Fig.~\ref{CO-grid}{\it a} overlaid with contours of the radio continuum emission from the SNR shock wave. 
Overall, the \element[][12]{CO} compared to the \element[][13]{CO} emission appears spatially more extended within the region of interest. 
This difference in distribution can be explained by the fact that \element[][13]{CO} emission is optically thinner than that from \element[][12]{CO}, whereby it gives account of more internal and denser regions in the cloud \citep{Wilson+13}. 
Notably, in the brightest part of the uncovered molecular structure, corresponding to the yellowish regions where \element[][12]{CO} and \element[][13]{CO} emissions overlap, {\kes} exhibits (within the 80{\asec} resolution of the radio image smoothed to match the resolution of the CO data) a significant deviation from a spherical symmetry. 

In order to gain further insight into the characteristics of the molecular gas, we extracted \element[][12]{CO} and \element[][13]{CO} spectra across the entire region where the radio continuum emission from, and the CO line emission towards, {\kes} show a 
line-of-sight superposition. 
To cover this region comprehensively, we employed a grid of \dms{1}{\prime}{5} boxes, as depicted in Fig.~\ref{CO-grid}{\it a}. 
To enhance the signal-to-noise ratio of the spectra, they were smoothed by averaging intensity values within the 5 nearest-neighbour velocity channels. 
The resulting spectra, presented in Fig.~\ref{CO-grid}{\it b}, show a broad emission region in \element[][12]{CO} with a velocity span $\Delta v \simeq$ 17~{\kms}  and an intensity varying from approximately 2 to 4~K. 
Multiple \element[][12]{CO} kinematic components contribute to the observed emission with peak velocities at approximately $-30$, $-25$, and $-20$~{\kms}. 
Examples of profiles showing this behaviour correspond to boxes 16-18, 22-24, and 28-30, all of them within the outermost radio contour of the SNR shell. At the easternmost border of {\kes}, these three velocity components appear to be less distinguishable and exhibit blending. 
We stress that the \element[][13]{CO} profiles (Fig.~\ref{CO-grid}{\it b}) do not reproduce the triple-peaked structure observed in the \element[][12]{CO} gas, but a broad peak at about $-20$~{\kms}, in complete agreement with the velocity of the \ion{H}{I} absorption features used to constrain the distance to {\kes} \citep{Ranasinghe+Leahy-22_distances}. The intensity of the \element[][13]{CO} emission peaks is $\sim$ 1-2~K.

Figure~\ref{CO-PV} displays position-velocity ($p$-$v$) diagrams of the molecular gas emission. They were constructed by integrating the \element[][12]{CO} emission along the R.A. direction in seven slices, each covering a range of 100{\asec}. These slices span the entire cloud of interest and constitute an appropriate tool to effectively capture the spatial heterogeneity of the individual velocity components observed in the spectral distribution shown in Fig.~\ref{CO-grid}{\it b}.  
By inspecting the $p$-$v$ diagrams, it is evident that the molecular emission is mostly concentrated at $-20$~km~s$^{-1}$, adjacent at the position where the radio shell of the remnant is highly distorted and it appears to branch off to the interior of the SNR (panels {\it e} to {\it g} in Fig.~\ref{CO-PV}). Bright knots are noticeable in the cloud's interior. 

By combining the CO emission and \ion{H}{I} absorption profiles (not shown here) extracted over the brightest part of the molecular concentration  emitting at $-20$~{\kms}, we determined that it is located at its far kinematical distance $\simeq 8$~kpc. The remaining velocities components of the cloud are at around 7~kpc ($-30$~{\kms}) and 7.5~kpc ($-25$~{\kms}). 
Taking the associated uncertainties ($\simeq$1.0~kpc) in these determinations into account, it can be concluded that these peaks arise from different components of the same structure. The average distance to this structure is estimated to be approximately 7.5~kpc, completely compatible with the distance determined for {\kes} ($\simeq 7.9$~kpc \citealt{Ranasinghe+Leahy-22_distances}). 
The error in the distance determination for the molecular gas stem from various factors. One of these contributions is the uncertainty in the peak velocity value for each gas component. 
Additionally, accurately measuring the properties of individual molecular components can be challenging, especially when they are not completely resolved. 
Lastly, the use of a Galactic rotation curve, such as the one proposed by  \citet{Reid+14}, involves assumptions and uncertainties. 

We notice that despite our analysis of the molecular material through \element[][12]{CO} and \element[][13]{CO} lines supports the coexistence of the discovered eastern cloud and the remnant, we have not observed distinct broadenings in the CO emission attributable to turbulence caused by the impact of {\kes}'s shock front. 
Therefore, we propose that the spectral behaviour of CO might illustrate a soft contact between the surrounding cloud and the remnant's shockwave. Certainly, the process of impacting the cloud might be at an initial stage.

We have also estimated the total mass $M$ and mean density $n(\mathrm{H}_2)$ of the molecular gas in the newly-detected cloud at $v_\mathrm{LSR}$ $\simeq$ $-31$ to $-14$~{\kms} by using both the \element[][12]{CO} and \element[][13]{CO} ($J$= 1-0) emissions. 
The procedure involves calculating the molecular hydrogen column density $N(\mathrm{H}_2)$, and deriving both $M$ and $n(\mathrm{H}_2)$ from it. 
$N(\mathrm{H}_2)$ is obtained from the integrated emission of the CO by using appropriate conversion factors relating the integrated emission of \element[][12]{CO} and the H$_2$ column density ($\mathrm{X_{12}} = 2.0 \times 10^{20}$~\cm{-2}~(K~{\kms})$^{-1}$, \citealt{Bolatto+13}), and also between the column density $N(^{13}\mathrm{CO})$ and $N(\mathrm{H}_2)$ ($\mathrm{X_{13}}$ = $7.7 \times 10^5$, \citealt{kohno+21}). 
We refer the reader to the work of \citet{Wilson+13}, where the expressions and assumptions (related to local thermodynamic equilibrium) employed for obtaining column densities are explained in detail. 
The mass is calculated through the relation $M = \mu\,m_\mathrm{H}\,\Omega\,D^2\,N(\mathrm{H}_2)$, where $\mu = 2.8$ is the mean molecular mass of the cloud,%
\footnote{Assuming a helium abundance of 25\%.} 
$m_\mathrm{H}$ is the hydrogen atom mass, and $\Omega$ is the solid angle subtended by the cloud located at the distance $D$. 
On the other side, the mean molecular density is $n(\mathrm{H}_2) = N(\mathrm{H}_2) / l$, where $l$ denotes the extent of the cloud in the line of sight assumed to be equal to the average of the mean size of the structure in R.A. and Dec. 
For the integration of the CO emissions we used a circular region with a radius of 5{\amin} (or $\sim 11$~pc at a distance of $\sim 7.5$~kpc to the cloud) centred at \RA{13}{06}{30}, \Dec{-62}{43}{20}. 
From this integration, we derived a molecular column density $N(\mathrm{H}_2) \approx 1 \times 10^{22}$~\cm{-2}, consistent for both \element[][12]{CO} and \element[][13]{CO} gases. Therefore, the 
resulting mean mass and  molecular density for the eastern cloud were estimated to be $M \approx 4.2 \times 10^4$~{\Msun} and  $n(\mathrm{H}_{2}) \approx 300$~\cm{-3}, respectively. 
The uncertainties in these measurements are of the order of 40\%, and comprise errors in the distance and the definition of the structure in the plane of the sky, as well as in the velocity space. 
We also notice that differences in the obtained values using emissions from both  \element[][12]{CO} and \element[][13]{CO} were found to be within 20\%. 
The fact that the values estimates from both \element[][12]{CO} and \element[][13]{CO} are in agreement indicates that both isotopologues provide consistent measurements and can be used for deriving cloud parameters. 

Now we focus on the molecular gas distribution towards the western side of the SNR shell. Of particular interest is the absence of CO emission above 3$\sigma$ (where $\sigma$ $\sim$ 1~K) spatially correlated with the radio continuum bright region, which is roughly 2{\amin}~$\times$~4{\amin} in size (centred at \RA{13}{05}{30}, \Dec{-62}{41}{10}). 
In this region, molecular hydrogen and ionic lines at infrared wavelengths have revealed the expansion of the SN shock on a molecular cloud, as reported by \citet{lee+19-FeII}. 
Based on these findings, we tentatively propose the existence of a ``CO-dark'' gas component to the west of {\kes}. In this scenario, the gas-phase carbon could be in atomic form, while the hydrogen is in molecular form. A similar phenomenon have been observed in CTB~37A \citep{Maxted+13} and RX~J1713.7--3946 \citep{Sano+Fukui-21}. 
More sensitive CO molecular-line measurements are needed to shed more light on this scenario for {\kes}. It is, however, worth noting that the detection of $\gamma$-ray radiation in the direction of the remnant can indirectly trace the dark-molecular gas component if it is generated by cosmic-ray collisions with the gas \citep{Wolfire+10}. 
In the case of {\kes}, a $\gamma$-ray excess at GeV energies has indeed been detected in projected coincidence with the remnant. The analysis of this emission is addressed in Sect.~\ref{gamma} of this work.  
In passing by, a peculiar wall-like structure of \element[][13]{CO} is observed at a distance of approximately \dms{1}{\prime}{5} from the outermost radio contour towards the west. However, the straight vertical border of this structure and the absence of a counterpart in the \element[][12]{CO} data covering the same region strongly suggest that this is not a  real feature.

\section{The field of {\kes} at $\gamma$-ray energies}
\label{gamma}
The first reports of emission in the $\gamma$-ray domain spatially projected onto  {\kes} were presented by \citet{Wu+11} and \citet{Gelfand+13}, based on statistics of 30-39 months data from {\Fermi}-LAT. 
The latest {\Fermi}-LAT catalogue of GeV sources (4FGL-DR3,%
\footnote{\url{https://fermi.gsfc.nasa.gov/ssc/data/access/lat/10yr_catalog/}.} 
\citealt{Abdollahi+20,Abdollahi+22_4FGL-DR3}) identifies the observed $\gamma$-ray excess directed towards {\kes} as 4GL~J1305.5$-$6241. To date, there have been no reports of TeV radiation detected in the {\kes}'s field. In this section we provide an update on the GeV emission in direction to this SNR and investigate the nature of the high energy photons by modelling the spectral energy distribution (SED) combining the updated $\gamma$-ray data with the firstly-obtained radio continuum spectrum of {\kes} presented in Sect.~\ref{radio-spectrum}.

\subsection{The treatment of GeV data from {\Fermi}-LAT}
Our analysis comprises the largest statistics of events for {\kes} to date, consisting of approximately 14.5~yr of continuous data acquisition with the {\Fermi}-LAT, spanning from the beginning of the mission on August $4^\mathrm{th}$, 2008, to February $24^\mathrm{th}$, 2023.\footnote{Corresponding to the time range from 239557417 to 681653590 seconds of the mission elapsed time (MET).} This represents a significant improvement of $\sim$450\% in observing time compared to the previous study conducted by \citet{Gelfand+13}.

The processing of the LAT data was conducted using the {\tt fermipy} module version 1.1.6 \citep{Wood+17_fermipy}, which uses the {\it Science Tools} package version 2.2.0.\footnote{{\tt fermipy} routines were implemented through a {\tt JupyterLab} Notebook, \url{https://jupyter.org/}.} 
Events were selected using the Pass~8, $3^\mathrm{rd}$ release ({\tt P8R3}) of photon reconstructions and the latest instrument response functions ({\tt P8R3\_SOURCE\_V6}). 
The region of interest (ROI) used to extract the events was a circle $15^\circ$ in size centred at {\kes} (\RA{13}{05}{53}, \Dec{-62}{42}{10}). 
To extract valid events, we employed the tasks {\it gtselect} and {\it gtmktime} applying standard filters for good-time intervals (GTIs)%
\footnote{Information about the definition of events and GTIs can be found in the {\Fermi}-LAT Science Support Center (FSSC) web page, \url{https://fermi.gsfc.nasa.gov/ssc/}.} and selecting ``source'' class events ({\tt evtype = 3}). 
An additional cut for the zenith angle at 90{\degree} was implemented to minimise potential contamination from cosmic-ray (CR) interactions in the upper atmosphere. 
We considered events with reconstructed energies above 0.3~GeV to mitigate the adverse effects of systematic uncertainties in the effective area and the degradation of the point spread function (PSF) at the lowest energies \citep{Ackermann+12_Fermi}. Furthermore, we excluded photons above 300~GeV due to the limited amount of events at the highest energies.

The set of filtered events was used to fit a sky model through a maximum likelihood optimisation procedure \citep{Mattox+96}. For the optimisation, we implemented a binned likelihood analysis over the ROI. The spatial bins were set at \dms{0}{\circ}{01} for both morphological and spectral analysis, and the energy range from 0.3 to 300~GeV was divided into 10 logarithmic bins per decade. 
In the analysis we included all sources from the LAT 10-year (4FGL-DR3) located within the ROI, except {\FGLsource} in projected coincidence with {\kes}. 
The models used for the optimisation were {\it gll\_iem\_v06} for the diffuse Galactic background and {\it iso\_P8R2\_SOURCE\_V6\_v06} for the isotropic background. The spectral parameters and normalisations of these models were allowed to freely vary during the optimisation process. 
Convergence was achieved by means of the {\tt Minuit} optimiser, fixing the source parameters beyond a radius of 4{\degree} from the ROI centre. 

\subsection{Morphological and spectral characteristics of the GeV emission}
\label{gamma-morph}
\begin{figure}[h!]
 \centering
 \includegraphics[width=0.47\textwidth]{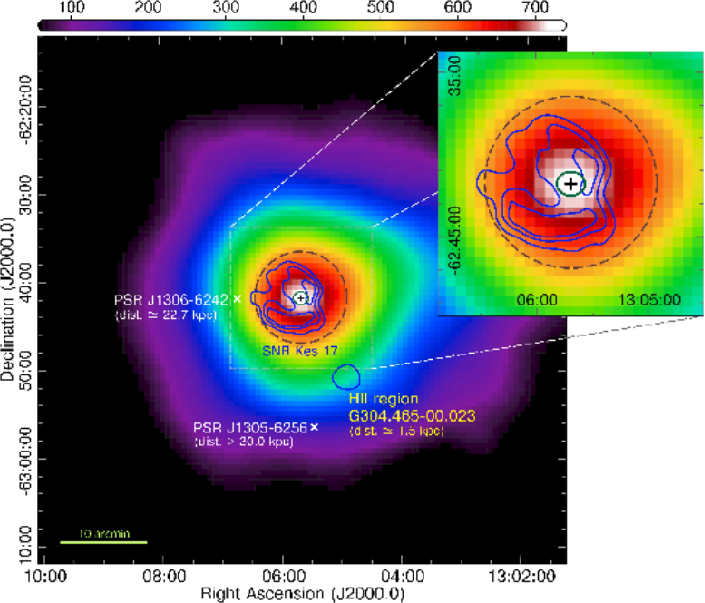}
 \caption{
 Test-statistics map with \dms{0}{\circ}{01} pixel size in the 0.3-300~GeV energy band in the {\kes} SNR's field (see text for details). 
 Blue contours delineating the SNR radio synchrotron emission are from the 843~MHz SUMSS map (levels: 0.015, 0.15,  and 0.30~{\Jyb}, at the resolution of 45$^{\prime\prime}$). 
 The positions and distances of the \ion{H}{II} region {\HII304} \citep{Urquhart+22} and the two known pulsars, PSR~J1306--6242 \citep{Kramer+03} and PSR~J1305--6256 \citep{Manchester+01}, lying in the field are also labelled. 
 The plus symbol and dark-green ellipse within the {\kes} radio contours mark respectively the best-fit position and 95\% confidence region for the GeV emission obtained with a point-source spatial template. The dashed brown circle traces the 95\%-confidence fitted size of the $\gamma$-ray source. 
 The inset shows a $\sim$15{\amin}~$\times$~15{\amin} close-up view of the {\kes} region, centred at the position of the $\gamma$-ray excess.}
 \label{TS-map}
\end{figure}

To investigate the spatial distribution of the $\gamma$-ray emission, we used the {\tt fermipy} tool {\it tsmap} to construct a test-statistics (TS) map. Each pixel in this map represents the likelihood of having a point source at the corresponding coordinates, compared to the null hypothesis of the sky model without the point source. 
The TS parameter is calculated as TS = $2\ln(L/L_0)$, where $L$ and $L_0$ are the likelihoods including and excluding the source in the sky model, respectively. The TS map allows us to assess the significance of the source detection, $\sigma \sim \sqrt{\mathrm{TS}}$ \citep{Mattox+96}. 
In Fig.~\ref{TS-map} we present the TS map obtained, overlaid with contours depicting the radio emission from {\kes}. 
The global TS value after the likelihood optimisation for the GeV source projected in coincidence with {\kes} was  TS $\simeq$ 730, corresponding to a detection significance $\sigma \simeq 27$. This represents an improvement of about 2.5 times over the value reported by \citet{Gelfand+13}.

We tested the possibility that the GeV emission be extended by using the {\it extension} tool from {\tt fermipy}, which makes a likelihood analysis by modelling the source as a circular region with variable radius. 
After convergence, the fitted value of the radius at 95\% confidence is \dms{0}{\circ}{09}, and the significance of the extension is TS $\simeq$ 16.3 ($\sigma \sim 4$). The low significance of the fitted size is indicative that the emission is not significantly extended. 
We then consider in the following that the $\gamma$-ray excess detected by the {\Fermi}-LAT corresponds to a point-like source and, consequently, we modelled it as a point source. Under this assumption, the {\it localise} tool yields the following location of this source: 
R.A. = \RA{13}{05}{40.71} $\pm$ $01.66^\mathrm{s}$, 
Dec. = \Dec{-62}{42}{01.6} $\pm$ 21.6{\asec}, 
within a 95\% confidence limit. 
From our analysis the localisation of the source, indicated by a green ellipse in Fig.~\ref{TS-map}, has been significantly improved by approximately one order of magnitude compared to the previous value reported in the 4FGL catalogue. 
The molecular gas, traced by the bright IR filaments (\citealt{Lee+11_FIR}, and references therein) could extend to cover the  $\gamma$-ray region. 
As we proposed in Sect.~\ref{CO}, this region is suspected of containing ``CO-dark'' molecular gas in interaction with {\kes}'s shock front. 

To investigate the spectral characteristics of the GeV $\gamma$-ray excess detected towards {\kes}, we generated an SED by performing a binned likelihood analysis in the 0.3-300~GeV range, implemented through the {\tt fermipy} tool {\it sed}. 
To ensure a balance between energy resolution and statistical significance, the data were grouped into 5 equally-spaced bins per decade on a logarithmic energy scale. 
Additionally, we incorporated energy dispersion corrections to mitigate systematic effects on the fitted spectral parameters. 

In our treatment, systematic contributions arise from the effective area, the PSF of the {\Fermi}-LAT, the energy scale, and variations in the spectral parameters due to the normalisation of the diffuse background.\footnote{Further details about systematic errors can be found in the FSSC web page.} 
The contribution related to the effective area is variable and can reach approximately 10\% at the extremes of the energy range considered in our analysis. For energies below 100~GeV, the systematic error related to the PSF containment radius is around 5\% and increases linearly to about 20\% for higher energies. 
Energy scale uncertainties are within 5\% throughout the energy range. 
On the other hand, systematics associated with the Galactic diffuse background were estimated following the procedure from \citet{Abdo+09_systematics}, which consists in artificially varying and fixing the normalisation by  $\pm$6\% with respect to the original fit and examining the resulting variations in the fitted spectral parameters. 
In the analysis of the broadband SED (Sect.~\ref{SED-analysis}), both statistical and systematic effects were considered. The contributions from both sources of uncertainties were added in quadrature to obtain a final error band for each energy bin.

\begin{figure}[h!]
 \centering
 \includegraphics[width=0.45\textwidth]{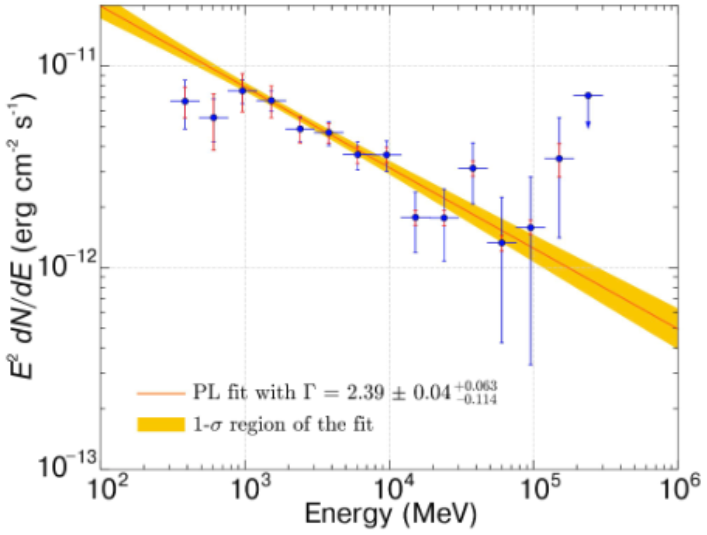}
 \caption{Spectral energy distribution of the $\gamma$-ray emission in the region of {\kes}, as detected by {\Fermi}-LAT. The orange line indicates the best fit to the data using the power-law model $dN/dE = {\phi_0}(E/E_0)^{-\Gamma}$ with $\Gamma= 2.39 \pm 0.04 \,^{+0.063}_{-0.114}$ (see text for details). 
 The yellow-shaded zone corresponds to the 1-$\sigma$ region of the spectral fit. 
 Blue and red error bars represent statistical and systematic uncertainties in the spectral points, respectively.}
 \label{SED-gamma}
\end{figure}

The data were fitted using a power-law model $dN/dE = {\phi_0}(E/E_0)^{-\Gamma}$, where $\phi_0$ is the differential flux (in units of \cm{-2}~s$^{-1}$~MeV), $E_0$ is the ``pivot'' energy, and $\Gamma$ represents the energy spectral index. 
Through the bin-by-bin likelihood procedure, we obtained a spectral index value $\Gamma= 2.39 \pm 0.04 \,^{+0.063}_{-0.114}$ ($\pm$ stat $\pm$ syst), which is well in agreement with the 4FGL value, but it is softer and marginally consistent with that from \citet{Gelfand+13}. 
Our result is also in agreement with the correlation observed by \citet{Acero+16} between the radio spectral index $\alpha$ and the GeV photon index $\Gamma$ for SNRs interacting with molecular clouds. 
The data points are shown in Fig.~\ref{SED-gamma}, where blue and red error bars denote the statistic and systematic uncertainties, respectively. 
The plot also shows the power-law fit to the data points, as well as the 1-$\sigma$ confidence interval for the fit. 
The integrated flux is determined to be $F(0.3\mathrm{-}300~\mathrm{GeV}) = (2.98 \pm 0.14) \times 10^{-11}$~erg~cm$^{-2}$~s$^{-1}$,%
\footnote{Equivalent to an integrated photon flux of $(1.87 \pm 0.09) \times 10^{-8}$~ph~\cm{-2}~s$^{-1}$.} 
corresponding to a luminosity $L_{\gamma}(0.3\mathrm{-}300~\mathrm{GeV}) = (2.22 \pm 0.45) \times 10^{35}$~erg~s$^{-1}$ at the distance of $7.9 \pm 0.6$~kpc. 
The phenomenology of how the luminosity in $\gamma$-rays  for SNRs competes with factors such as distance uncertainties, molecular gas repository densities, or time evolution effects is complex and a detailed analysis of this topic is beyond the scope of this work. 
However, despite this limitation, we can provide a brief comparison of our $L_{\gamma}$ estimate for {\kes} with those derived in a similar energy range (0.1-100~GeV) for emitters identified as advanced (i.e., from middle-aged to old, $\gtrsim 10$~kyr) and young-aged ($\lesssim 3$~kyr) SNRs associated with molecular clouds (for a more detailed discussion, refer to \citealt{Acero+22_Kepler}). 
For instance, when considering  the older sources W44 and IC~443 with high local densities ($\sim 10^2$-$10^4$~cm$^{-3}$ \citealt{Yoshiike2013,DellOva2020}), they would be 5 or even 60 times more luminous than  {\kes} if we place them at the distance of $\sim$8~kpc.%
\footnote{The reported $\gamma$-ray luminosities values turns out to be $10^{34}$-$10^{35}$~erg~s$^{-1}$ in the 0.1-100~GeV range for W44 and IC~443 according to \citet{Acero+22_Kepler} 
($d_{\mathrm{W44}}\simeq 3$~kpc, \citealt{Ranasinghe+Leahy-22_distances} and $d_{\mathrm{IC~443}}\simeq 1.7$~kpc, \citealt{Yu2019}).}. 
Additionally, the $\gamma$-ray luminosity of the mature remnant Cygnus Loop, evolving in a low-density environment ($\sim$1-10~cm$^{-3}$, \citealt{Fesen2018}), would be comparable to our estimate for {\kes} if they were located at the same distance.%
\footnote{The estimated luminosity in 0.1-100~GeV for Cygnus Loop is $\simeq 10^{33}$~erg~s$^{-1}$ \citep{Acero+22_Kepler} at 0.7~kpc \citep{Fesen2018}.} 
We also point out that  all of these middle and advanced-age remnants are more luminous, by one to two orders of magnitude, compared to the young SNRs Tycho and Kepler, which are expanding in low-density media 
($\sim$10~cm$^{-3}$, \citealt{Acero+22_Kepler,Zhang2013}).%
\footnote{$L_{\gamma}$ for Tycho and Kepler spans the range $\simeq10^{33}$-$10^{34}$~erg~s$^{-1}$ at 4 and 5~kpc, respectively (\citealt{Acero+22_Kepler}, and references therein).}

\subsection{Analysis of the spectral energy distribution of {\kes} from radio to $\gamma$ rays}
\label{SED-analysis}
In this section we study the spectral energy distribution of {\kes}, incorporating the new nonthermal continuum radio spectrum extracted from the SNR shell and the high-energy spectrum obtained from new observations by the {\Fermi}-LAT. 
For the nonthermal X-ray emission, we used an upper limit derived from \it Suzaku \rm data by \citet{Gelfand+13}. At very high energy, we used an upper limit above 1~TeV from the H.E.S.S. Galactic Plane Survey \citep{Gangoso-14,Abdalla+18}. 
To model the multiwavelength emission, we considered an electron population that produces synchrotron radiation, inverse Compton scattering (IC), and nonthermal bremsstrahlung. Additionally, we incorporated a proton population that interacts with the surrounding gas, resulting in the subsequent production and decay of neutral pions ($\pi^{0}$). 
The parameters characterising these models are presented in Table~\ref{gamma-table} and were derived using the {\tt Naima} Python package \citep{Naima}. 

\begin{figure}[h!]
 \centering
 \includegraphics[width=0.47\textwidth]{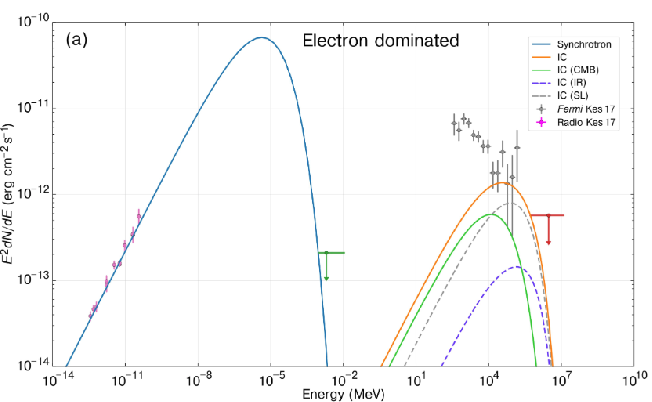}
 \includegraphics[width=0.47\textwidth]{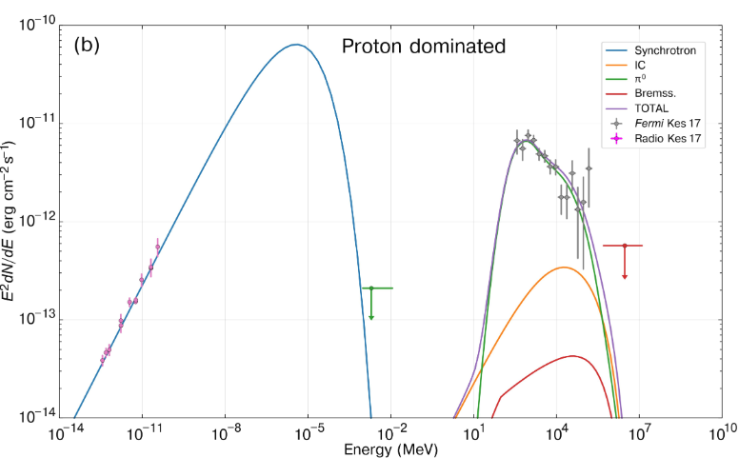}
 \caption{Broadband SED from radio to $\gamma$ rays for {\kes}. Fluxes in the radio band correspond to those in Table~\ref{fluxes-table} (in units of erg~\cm{-2}~s$^{-1}$), while those at $\gamma$-ray energies correspond to the new {\Fermi}-LAT data from this work (Sect.~\ref{gamma-morph}). 
 Upper limits at X-ray and TeV energies where derived from the non detection of X-ray synchrotron emission and the H.E.S.S. observatory, respectively (see text for details). 
 ({\it a}) SED modelling with a unique electron population producing synchrotron and IC radiation over the broadband spectrum. 
 ({\it b}) SED modelling with an ExpCut-off proton-dominated model producing synchrotron, IC, and Bremsstrahlung radiation from a parent electron population, and $\gamma$-ray emission via the decay of $\pi^0$ created from a proton population interacting with the surrounding gas (see Table~\ref{gamma-table} for a complete list of parameters).} 
 \label{SED-radio-gamma}
\end{figure}

To investigate the plausibility of a scenario where the leptonic component is dominant, we considered a one-zone model with an electron population distributed in energies according to a power-law. 
The updated radio continuum spectrum, extended to cover frequencies from 88 to 8800~MHz, allowed us to further constrain the spectral index of the synchrotron emission, at $\alpha = -0.488 \pm 0.023$ (Sect.~\ref{radio-spectrum}). We used this measurement to fix the initial power-law index $\Gamma_\mathrm{e} = 1 - 2\alpha$ of the electron energy spectrum to a value $\approx$ 1.9. 
Then, the X-ray upper limit derived by \citet{Gelfand+13} imposes a tight constrain on the magnetic field ($B$), the energy cut-off ($E_\mathrm{c}$) and the energy density ($W_\mathrm{e}$) of the electron population, which are degenerate. 
Following \citet{Gelfand+13}, we assumed a magnetic field $B = 35~\mu$G, which yields a maximum value of the cut-off energy estimated to be $E_\mathrm{c} \approx 2 $~TeV and a total energy density $W_\mathrm{e} = 4.3 \times 10^{48} (d_\mathrm{SNR}/7.9~\mathrm{kpc})^2$~erg, which appears reasonable for a middle-aged system (see discussion in \citealt{Gelfand+13} about the cut-off in the spectrum due to synchrotron losses of middle-aged to old systems). 
We first consider the electron population distribution obtained from the radio spectrum synchrotron fit  to compute the associated Inverse Compton emission in order to reproduce the observed level of measured $\gamma$-ray emission. We consider three interstellar radiation fields: 
the cosmic microwave background (CMB) ($T_\mathrm{CMB}$ = 2.72~K, $u_\mathrm{CMB}$ = 0.26~eV~cm$^{-3}$), the far-IR (FIR) radiation ($T_\mathrm{FIR}$ = 27~K, $u_\mathrm{FIR}$ = 0.415~eV~cm$^{-3}$), and the near-IR starlight radiation  ($T_\mathrm{SL}$ = 2800~K, $u_\mathrm{SL}$ = 0.8~eV~cm$^{-3}$), computed from the {\tt GALPROP}%
\footnote{\url{https://galprop.stanford.edu/index.php}} 
model at the position of the remnant (Galactocentric distance of $\sim$6~kpc) \citep{Strong+04,Porter+06}. 
However, as shown in Fig.~\ref{SED-radio-gamma}, the IC radiation produced by this electron population fails to reproduce the new {\Fermi}-LAT spectrum. Particularly, the shape of the GeV emission cannot be reproduced by a simple electron population with an index of 1.9. Furthermore, the upper limit from H.E.S.S. places strong constraints on the level of IC emission at very high energies, and constrains the maximum value of the energy break in the electron spectrum to be 1.5~TeV and sets a minimum value for the magnetic field strength at 35~$\mu$G (see Table~\ref{gamma-table}). 
Therefore, we conclude that it is not possible to adequately model the broad-band emission using a purely leptonic scenario, at least within the framework of a one-zone model.

\begin{table*}[h!]
\small\centering
-------------------------------------
\caption{Parameters results from the two models used to reproduce the broadband SED of Kes~17 (see Fig.~\ref{SED-radio-gamma}). One model considers a unique electron population, and the other one is a proton-dominated model that includes both electrons and protons. $B$ is the magnetic field at the shock, $n_0$ is the average density of the surrounding ISM, while $\Gamma_\mathrm{e}$, $\Gamma_\mathrm{p}$, $E_\mathrm{b,e}$,  and $E_\mathrm{b,p}$ are respectively the indices and the energy cut-off or break in the parent electrons and protons spectra (see text for details). $E_\mathrm{min,e}$ and $E_\mathrm{min,p}$ are the minimum energy of the electron and proton spectra, whereas $W_\mathrm{e}$ and $W_\mathrm{p}$ are the total energy density of accelerated electrons and protons, respectively.}
\begin{tabular}{lccccccccccc}
\hline\hline
     \multirow{2}{*}{~~~~~~~~~~~~~~~~Scenario}                &  $B$     & $n_0$    & $\Gamma_\mathrm{e}$ & $E_\mathrm{min,e}$ & $E_\mathrm{b,e}$ & $\Gamma_\mathrm{p}$ & $\Gamma_\mathrm{p,1}$/$\Gamma_\mathrm{p,2}$ & $E_\mathrm{b,p}$ & $E_\mathrm{min,p}$ &  $W_\mathrm{e}$      & $W_\mathrm{p}$     \\
                             & ($\mu$G) & (cm$^{-3})$ &            & (MeV)  &  (TeV)           &            &                           &  (TeV)           & (MeV)  & (erg)    & (erg)  \\
\hline
Electron dominated           &  35  &           & 1.9        &  10    &  1.5             &            &                           &                   &  10     & $4.3\times10^{48}$ &        \\
\hline
Proton dominated  (BPW)      &  70  & 400       & 1.9        &  10    &  1.0             &            & 2.4/3.7                   &   2               &  10     & $1.5\times10^{48}$ &  $2.97\times10^{49}$ \\
Proton dominated (ExpCut-off)&  70  & 400       & 1.9        &  10    &  1.0             & 2.4        &                           &   2               &  10     & $1.5\times10^{48}$ &  $2.97\times10^{49}$ \\
\hline
\label{gamma-table}
\end{tabular}
\end{table*}

We now investigate a scenario where the major part of the $\gamma$-ray emission is attributed to $\pi^0$ decay. At this stage, it is important to recall that the {\Fermi} source is spatially coincident with a part of the shell which has bright radio emission and exhibits IR filaments. 
While no molecular structures were detected in the region of bright $\gamma$-rays (Sect.~\ref{CO}), the study of molecular hydrogen and ionic lines, made it possible to highlight a signature of a shock due to an interaction of the SNR shell with a cloud \citep{Lee+11_FIR}. 
Consequently, in our analysis, we  consider only the molecular gas in the western region of {\kes}, as the dominant contributor to the GeV photon flux through hadronic interactions. 
In the following we will consider a cloud density of 400~cm$^{-3}$ in accord with the estimation made by \citet{Lee+11_FIR} from the IR emission. This value is an order of magnitude higher than the $\sim 10$~cm$^{-3}$ used in \citet{Gelfand+13}. 
To analyse the hadronic origin of the $\gamma$-ray radiation, we have assumed two models: a power-law proton spectrum with an index $\sim$ 2.4 and a cut-off above a few TeV (ExpCut-off), and a  broken power-law (BPW) with indices $\Gamma_\mathrm{p,1}$ = 2.40 and $\Gamma_\mathrm{p,2}$ = 3.5 below and above $E_\mathrm{b}$ = 2~TeV, up to a maximum energy of 10~TeV, which is strongly constrained by the steep spectrum deduced from {\Fermi} data above few tens of GeV and the non detection of {\kes} by H.E.S.S. 
We report all-proton dominated model parameters in Table~\ref{gamma-table}. 
Both models fit the data equally well. To simplify the presentation, we have plotted only the ExpCut-off model in Fig.~\ref{SED-radio-gamma}, which clearly shows that the $\gamma$-ray spectrum strongly supports the hadronic origin of the radiation. 
Since the hadronic $\gamma$-ray emission is proportional to the product of the kinetic energy in protons and the density of the medium, these parameters are closely correlated. 
Assuming that the total mass of the molecular cloud acts as the target material, we derived a total energy of cosmic-ray protons 
$W_\mathrm{p} = 2.97 \times 10^{49} (n_\mathrm{p}/400~\mathrm{cm}^{-3})^{-1}(d_\mathrm{SNR}/7.9~\mathrm{kpc})^2$~erg. 
As can be appreciated in Fig.~\ref{SED-radio-gamma}, the first two data points at the lowest $\gamma$-ray energies seem to deviate from a pure power-law shape  and appear to be compatible with the so-called ``pion-bump'' feature observed below a few hundred MeV. 
As discussed in \citet{Tang-18}, the combination of this rising feature in the spectrum with a steep spectrum beyond a few tens of GeV, is widely recognised as a characteristic signature of $\pi^0$-decay,   illuminating hadronic emission in SNRs. 
Such a $\pi^0$ signature is particularly observed in the growing class of advanced-aged GeV emitters SNRs
interacting with MCs, as W44 \citep{Giuliani+11}, IC~443 \citep{Ackermann+13_pion}, and W51C \citep{Jogler+Funk-16}. 
On the other hand, from the modelling of the data depicted in Fig.~\ref{SED-radio-gamma}, it is evident that the contribution of bremsstrahlung radiation from the previously considered electron population, with a similar density, is clearly a minor component in the GeV energy range and its spectral behaviour does not accurately reproduce the spectrum shape. 
In the final proton-dominated model we include this marginal contribution, as well as that from IC emission of an electron population with a total energy density of $W_\mathrm{e} = 1.5 \times 10^{48} (d_\mathrm{SNR}/7.9~\mathrm{kpc})^2$~erg. 
The $W_\mathrm{e}/W_\mathrm{p}$ ratio is 0.05 (see Table~\ref{gamma-table}), larger than that in the proton-dominated scenario (0.01) discussed by \citet{Gelfand+13}, but significantly smaller than their IC dominated scenario (0.1). It can be attributed to the reduction of $W_\mathrm{p}$ as a consequence of the higher density of target matter considered here. 
On the basis of the new measurements of the GeV and radio spectra, we conclude that although a single electron population is considered to reproduce the overall synchrotron emission of the remnant, the $\pi^0$ decay process may be primarily responsible for the point-like $\gamma$-ray emission detected 
towards {\kes} in the GeV energy range.

\section{Summary and Conclusions}
\label{summary}
Based on our comprehensive update on the radio and $\gamma$-ray radiations from {\kes}, along with our analysis of the molecular environment brightening in the \element[][12]{CO} and \element[][13]{CO}~($J$ = 1-0) lines, we have arrived at the following picture: 

\noindent
1- {\kes} was created in a stellar explosion that occurred approximately 11~kyr ago in an ambient medium with a density roughly 7~cm$^{-3}$. 

\smallskip
\noindent
2- The observed spectral shape of the shock front in {\kes} emitting  from 88 to 8800~MHz is adequately  fitted with a simple power law model of index $\alpha = -0.488 \pm 0.023$. 
The available radio data suggest that there is not ionised gas located in, around, or anywhere along the sightline to {\kes} that significantly impacts the integrated spectrum. If present, this ionised gas may produce an spectral curvature below 100~MHz due to free-free thermal absorption. 
More sensitivity and resolution low frequency radio data are key to spatially resolving spectral curvatures due to intrinsic properties related to the shock in {\kes} and its interaction with the immediate SNR's surroundings observed in CO and infrared lines. 

\smallskip
\noindent
3- The eastern part of {\kes} is wrapping a CO cloud. The main evidences of such an interaction include the distortion of the SNR shock and the distance to the CO cloud that is found to be completely compatible with the distance to the remnant. 
The average mass and density of this cloud are determined to be $4.2 \times 10^4$~{\Msun} and 300~\cm{-3}, respectively. 
Noteworthy, there is not appreciable CO emission detected towards the western region of the radio shell, where molecular hydrogen has been proven to be shocked by the {\kes}'s shock front. This suggests the presence of a CO-dark molecular gas. 

\smallskip
\noindent
4- No features of atomic hydrogen physically connected to {\kes} are detected at the sensitivity and resolution of the SGPS data used in this work. 

\smallskip
\noindent
5- In its evolution, {\kes} produces $\gamma$-ray photons at GeV energies, which have been observed by the {\Fermi}-LAT telescope. The flux and luminosity at 7.9~kpc in the 0.3 - 300~GeV energy band are estimated to be $(2.98 \pm 0.14) \times 10^{-11}$~erg~cm$^{-2}$~s$^{-1}$ and $(2.22 \pm 0.45) \times 10^{35}$~erg~s$^{-1}$, respectively. The spectra  of this high-energy emission has an index $\Gamma= 2.39 \pm 0.04 \,^{+0.063}_{-0.114}$. 

\smallskip
\noindent
6- Based on observational evidence and modelling of the broadband SED ranging from radio to $\gamma$ rays, it has been determined that a purely leptonic (IC) scenario is not favoured as an explanation for the emission from {\kes} observed at  GeV energies. Instead, the evidence suggests that the primary contribution to the $\gamma$-ray flux originates from the collision between the western part of the SNR shock front and a dense IR emitting region. Consequently, our analysis adds {\kes} to the list of SNRs whose emission at GeV energies is hadronic-dominated. 
$\gamma$-ray luminosities measured in this class of remnants exhibit differences that can be interpreted in terms of the amount of molecular gas, which serves as target material for cosmic-ray interactions, as well as time-evolution effects. 
Future observations conducted using state-of-the-art  instruments operating at the highest energies of the electromagnetic spectrum, coupled with improved resolution and sensitivity,  will contribute to refining our understanding of the spectral and morphological behaviour of {\kes} in the $\gamma$-ray regime.

\begin{acknowledgements}
We are very grateful for the thorough corrections provided by the anonymous referee. 
G. Castelletti and L. Supan are members of the {\it Ca\-rre\-ra del Investigador Cient{\'i}fico} of CONICET, Argentina. This work was supported by the ANPCyT (Argentina) research project with number BID PICT 2017-3320. 
This work and collaboration is also supported by the International Emerging Actions program from CNRS (France).
\end{acknowledgements}

\bibliographystyle{aa}
 \bibliography{biblio-kes17}
\end{document}